\documentstyle[epsfig,a4,12pt]{article}
\renewcommand{\baselinestretch}{1.2}
\epsfverbosetrue

\def\nuebar{\rm{\bar{\nu_e}}}
\def\nue{\rm{\nu_e}}
\def\th232{\rm{ ^{232} Th }}
\def\u235{\rm{ ^{235} U }}
\def\u238{\rm{ ^{238} U }}
\def\k40{\rm{ ^{40} K }}

\def\cs133{\rm{ ^{133} Cs }}
\def\i127{\rm{ ^{127} I }}
\def\li7{\rm{ ^{7} Li }}
\def\b2o3{\rm{B_2 O_3}}
\def\gso{\rm{Gd_2 Si O_5 (Ce)}}
\def\gd160{\rm{^{160} Gd}}

\begin{document}
\hfill AS-TEXONO/00-04 \\
\hspace*{1cm} \hfill February 14, 2001

\begin{center}
\large
\bf{
Measurement of Intrinsic Radioactivity in a GSO Crystal\\
}
\vspace*{0.5cm}
\large
S.C.~Wang~$^{a,}$\footnote{Present Address:
TRIUMF, Canada.},
H.T. Wong~$^{a,}$\footnote{Corresponding~author:
Email:~htwong@phys.sinica.edu.tw;
Tel:+886-2-2789-9682;
FAX:+886-2-2788-9828.},
M.~Fujiwara~$^{b,c}$\\
\end{center}

\normalsize

\begin{flushleft}
$^a$ Institute of Physics, Academia Sinica, Taipei 11529, Taiwan ROC.\\
$^b$ Research Center for Nuclear Physics, Osaka University, Ibaraki,
Osaka 567-0047, Japan.\\
$^c$Advanced Science Research Center, JAERI, Tokai, Ibaraki,
319-1195, Japan.
\end{flushleft}

\begin{center}
{\bf
Abstract
}
\end{center}

Scintillating crystal detectors offer
potential advantages in low-energy low-background
experiments for particle physics and astrophysics.
The GSO crystal is an interesting detector to
explore for future neutrino physics experiments.
The contributions 
to background due to the various channels of 
intrinsic radio-isotopes from the
$\th232$ and $\u238$ series are identified
and studied 
with time-correlation analysis and detailed fits
to the spectral shape.
Good agreement is achieved between
measured and simulated spectra, indicating
background suppression factors to the
$10^{-2} - 10^{-3}$ level are possible.
The procedures can be adopted for
background understanding and suppression in
other low-count-rate experiments where the
dominant source of background is from
internal radioactivity.
Based on 1656~hours of data taking,
limits on  the double beta decay half-life
for the various channels in $^{160}$Gd
are derived.
The limits for
the neutrinoless and the Majoron modes  are
$\rm{T^{0 \nu \beta \beta}_{\frac{1}{2}} 
> 2.0(0.8)  \times 10^{20}~y}$
and 
$\rm{
T^{ \chi \beta \beta}_{\frac{1}{2}} ~ > ~
3.0 (1.9) \times 10^{19}~y}$,
respectively,
at 68(90)\% confidence level.

\vspace*{0.1cm}

\begin{flushleft}
\end{flushleft}

\vspace*{1cm}

\begin{center}
{\it
Nucl. Instrum. Methods. A 479, 498 (2002).
}
\end{center}

\clearpage

\section{Introduction} 
\label{sect::intro}

Crystal scintillators have been widely used 
as radiation detectors as well as in medical and security
imaging~\cite{crystal}.
Large detector systems, of scale several tens of tons in mass,
have been constructed and made operational as
electromagnetic calorimeters in
high energy physics experiments~\cite{emcalo}. 
It is one of the few detector technologies (the other
being loaded liquid scintillator) where 
a large range of nuclei can be turned
into massive detector systems.

The merits of scintillating crystal detectors 
in non-accelerator particle physics experiments
have been recently discussed~\cite{prospects}.
Primarily, the high-Z composition of most crystals allows
a compact design and provides large suppression of
background due to ambient radioactivity if 
a three dimensional fiducial volume definition
can be realized. 
Experiment based on 100~kg of NaI(Tl) has been
built for Dark Matter searches, producing some
of the most sensitive results~\cite{dama}.
Another experiment toward 500~kg of CsI(Tl) for
studying low-energy neutrino interactions using
reactor neutrinos as probe is now under
construction~\cite{expt}.

An interesting isotope for low-energy  neutrino
physics experiment is $\gd160$, which has
an isotopic abundance of 21.9\% in natural gadolinium.
It is a double beta decay candidate~\cite{bbreview}. 
The lepton-number-violating neutrinoless double
beta decay ($0 \nu \beta \beta$)
channel, which is sensitive
to the Majorana neutrino mass terms,  proceeds as:
\begin{displaymath}
\rm{
^{160}Gd ~~ \rightarrow ~~
^{160}Dy ~~ + ~~ 2 ~ e^- ~~~~~~
( ~ Q ~ = ~ 1.73~MeV ~ ) ~~~.
}
\end{displaymath}
In addition, this isotope 
has recently been proposed as an attractive
isotope for the detection of low-energy solar
neutrinos~\cite{snureview} by providing a time-delay
signature for background suppression and for
tagging the flavor-specific $\nue$N-CC reactions~\cite{lens}:
\begin{displaymath}
\rm{
\nu_{e} ~ + ~ ^{160} Gd ~ \rightarrow ~
e^- ~ + ~ \gamma 's (prompt) ~ + ~ ^{160} Tb^* ~~~~~
(Q=169~keV) ~~ ,
}
\end{displaymath}
followed by, in the 100~ns time-scale,
\begin{displaymath}
\rm{
^{160} Tb^* ~ \rightarrow ~ ^{160} Tb ~
+ ~ \gamma ~~~~~~
(\tau_{\frac{1}{2}}=60~ns ~;~ E_{\gamma}= 63.7~keV) ~~.
}
\end{displaymath}

The scintillating crystal GSO ($\gso$), where
gadolinium is 74.4\% by weight, 
was first developed in the 1980's~\cite{gsofirst}.
It is an attractive technology to realize an
active detector based on gadolinium.
Various studies have been performed at the
prototype~\cite{gsoprop}  and detector~\cite{gsodet}
levels for calorimeter systems, as well
as for double beta decay searches~\cite{bbgso1,bbgso2,bbgso3}
and feasibility studies of solar neutrino
experiments~\cite{snugso}.

The properties of GSO, 
together with a few commonly used
crystal scintillators, is listed in Table~\ref{scincrystal}.
It can be seen that besides neutrino physics,
GSO may offer an attractive alternative
to BGO in high energy physics and medical applications,
having almost the same radiation length, but
higher light yield, faster decay constant, and
excellent radiation resistance.

In this article, we report on the measurement and analysis
of the major components of intrinsic radio-impurities  
in a commercially available 
GSO crystal\footnote{Manufacturer: Hitachi Co. Ltd., Japan.}.
These results are important to assess the feasibilities of
GSO crystals in the low-event-rate experiments mentioned,
and can serve as basis of future R\&D efforts.
In addition, the procedures can be adopted for
background understanding and suppression in 
experiments dominated by internal radioactivity
background. Lifetime limits on the various
double beta decay channels for $^{160}$Gd are derived.

\section{Experimental Setup}

\begin{figure}
{\bf (a)}
\centerline{
\epsfig{file=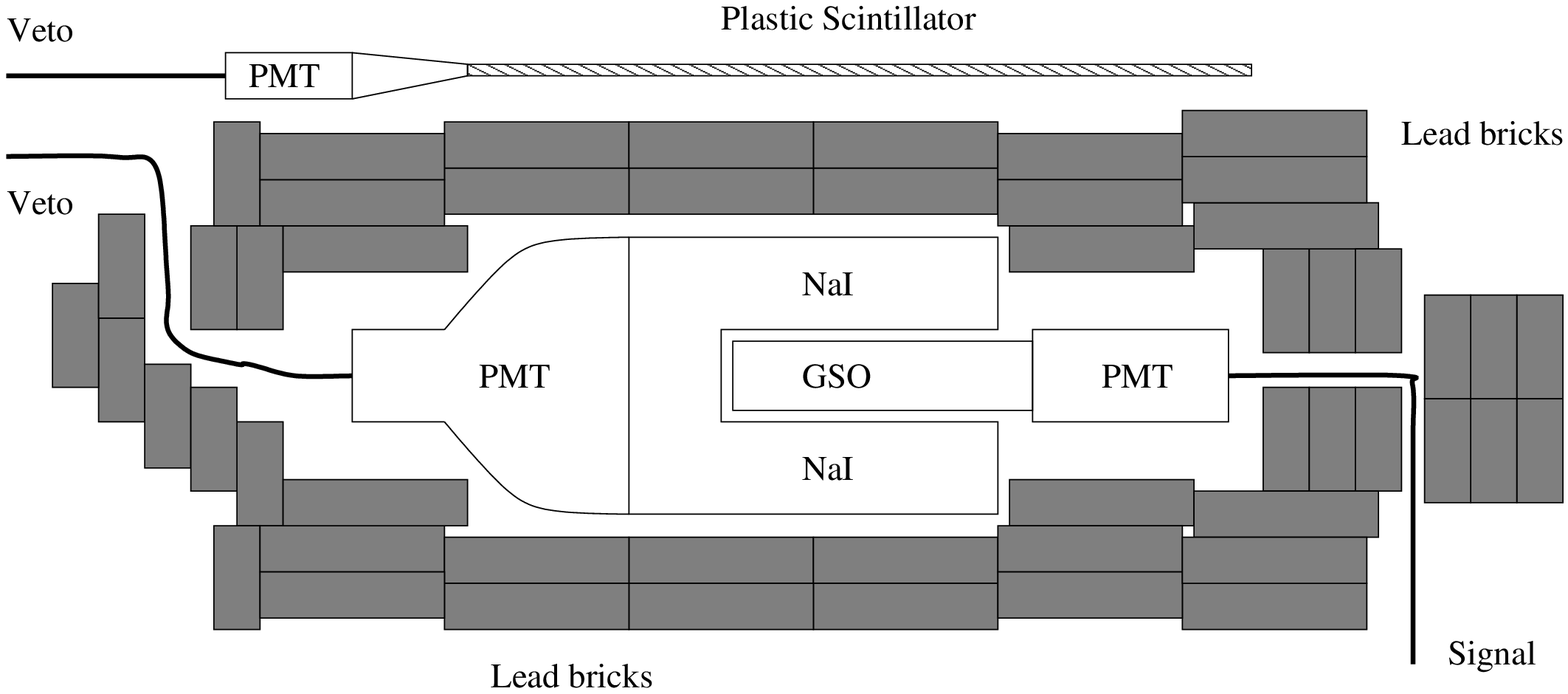,width=15cm}
}
{\bf (b)}
\centerline{
\epsfig{file=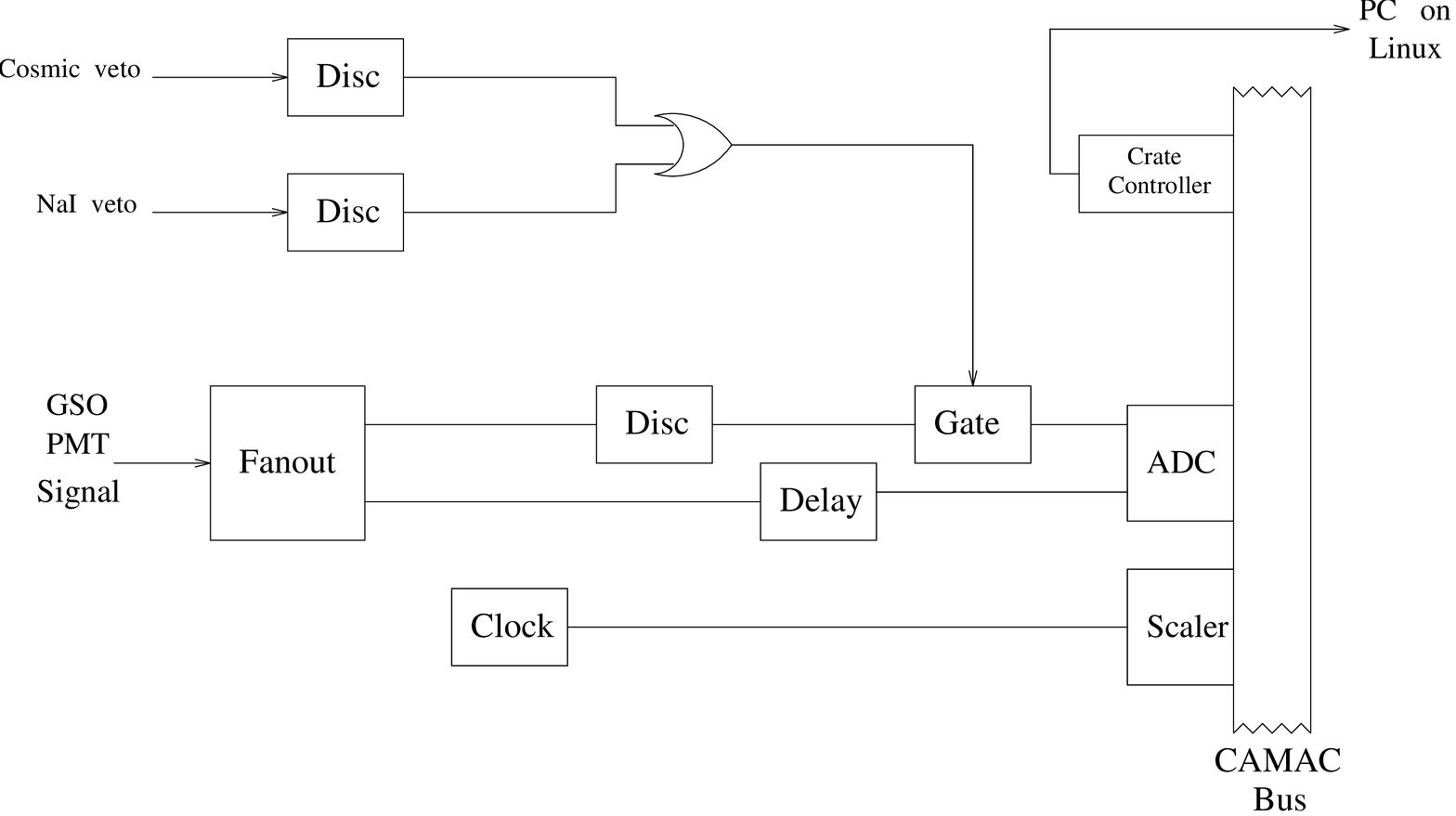,width=7cm,angle=270}
}
\caption{
Schematic drawing of (a) the experimental setup showing an
active NaI(Tl) veto scintillator around the GSO crystal,
together with a plastic scintillator
for cosmic-ray veto on top and  
4 $\pi$ lead shielding with at least 10~cm in thickness  ;
(b) the electronics and
data acquisition system.
}
\label{fig1}
\end{figure}

The schematic diagram of the experimental setup
is depicted in Figure~\ref{fig1}a.
The GSO crystal under study
has a dimension of 3.8 $\times$ 3.8 $\times$ 18 cm$^3$, weighing 1.744 kg
and was wrapped with white Teflon tapes to improve the light
collection. An outer layer of black vinyl tapes 
provided a light-proof layer.

A 2-inch diameter photo-multiplier(PMT) 
with potassium-free glass
was coupled to
one of the end surfaces of the crystal.
The whole crystal-PMT system was inserted into a well-shaped 
NaI(Tl) crystal scintillator\footnote{Manufacturer: Canbera}
with a mass of 14~kg, and viewed via a quartz window
by a 5-inch photo-multiplier tube with potassium-free glass.
The NaI(Tl) scintillator
served as an active shielding device, vetoing
background $\beta$'s and  $\gamma$'s from the ambient environment,
as well as the cosmic muons that penetrated the crystal detector. 
Surrounding the NaI-GSO-PMT active components was
lead shieldings of 10 to 15 cm in thickness. 
The measurement was performed in a ground-level
laboratory of a seven-storey building. The
estimated overburden is about 12~meter-water-equivalent,
sufficient to suppress the hadronic components of the
cosmic-rays.
A 40 $\times$ 50 $\times$ 1 cm$^3$ scintillation plastic was placed on
top of the detector outside the lead bricks for cosmic-ray muon vetoes.
As shown in Section~\ref{sect::results}, 
internal radioactivity becomes the dominant contribution to
the background spectrum under this shielding configuration.
Therefore, it is not necessary to make special additional efforts
to suppress the neutron and radon induced background in
the present work.

The schematic diagram of the electronics and
data acquisition systems is displayed in Figure~\ref{fig1}b.
The CAMAC system was adopted.
The energy and time for each event
were recorded.
The energy measurement was made
by using an analogue-to-digital 
converter\footnote{LeCroy 2249W} 
which integrated the anode current of the PMT
for a gate-time of 600~ns, 
while the timing measurement was done by a
scaler\footnote{Kinetic Systems 3615}
counting on a 1~MHz clock.
The scaler was read and then reset when
the LAM (Look-At-Me) line was set active by the ADC.
The timing accuracy was checked to be
better than the time bin of 1~$\mu$s.
Data were acquired and transferred 
through the CAMAC bus
via  a fast CAMAC Controller~\footnote{CAEN C111A}
to an interface processor
running on the Linux operating system with
a PC computer.
The rate at a detector threshold
of 50~keV after active vetoes and passive shieldings was 6.46 Hz.
The dead-time of the system, which was mainly due to the conversion
time of the CAMAC ADC (about 106 $\mu$s per conversion), 
is less than 0.1\%

\section{Results and Analysis}
\label{sect::results}

\begin{figure}
\centerline{
\epsfig{file=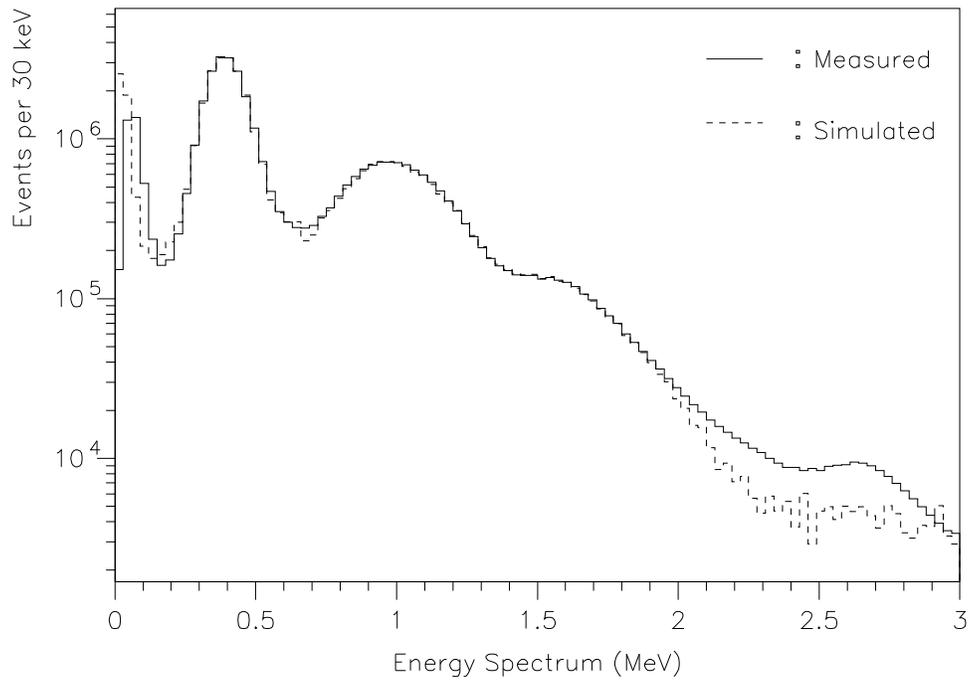,width=15cm}
}
\caption{
Comparison between the measured spectra from
1656 hours of data with the 
simulated spectra based on the best-fit rates
from the various components.
}
\label{fig2}
\end{figure}

Events from the GSO crystal not in coincidence
with the NaI(Tl) anti-Compton detector and 
the cosmic-ray veto panel were selected.
The energy spectrum
based on 1656~hours
of data taking is shown in Figure~\ref{fig2}. 
It can be seen that there are three distinct regions
in the energy spectrum due to intrinsic
radioactivity. 
The sharp peak at 0.394~MeV is due to the $\alpha$-decay
of the isotope $^{152}$Gd
in the crystal. The nominal isotopic
abundance of $^{152}$Gd is 0.2\%,
decaying by $\alpha$-emission at 2.14~ MeV and with 
a half life of
$1.08\times 10^{14}$~years~\cite{isotab}.
Events in the broad peak around 1 MeV
are due to the $\alpha$-decays of the radio-nuclides in the
$\th232$ and $\u238$
series in the GSO crystal. 
These isotopes and their
decay-daughters are primordial in origin
and get into the crystal 
mostly through the contaminations in the raw materials.
The high energy shoulder above 1.4 MeV and
up to 2.0 MeV is due to the $\beta$-emissions
together with the subsequent prompt $\gamma$-rays
from the radio-nuclides of 
these cascades.
The bump at the 2.6~MeV region is due to
the $^{208}$Tl 
2.61~MeV $\gamma$-rays from ambient radioactivity. 

There is no notice-able peak at 1.46~MeV due to the
$\gamma$-rays from the commonly found $^{40}$K,
since a PMT with potassium-free glass was used. In the
first trial measurement using another PMT with normal
glass, a peak at this energy did 
show up prominently.

\begin{figure}
\centerline{
\epsfig{file=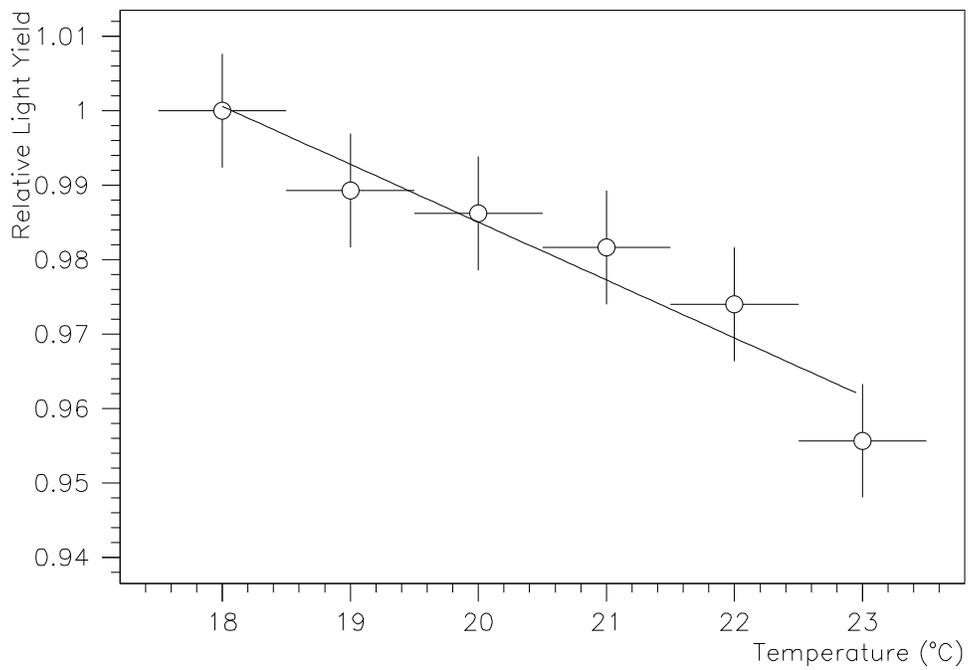,width=15cm}
}
\caption{
The measured light yield variation of the
GSO crystal, using
the $^{152}$Gd $\alpha$-peak as reference,
as a function of temperature
during the course of 1656~hours of data taking.
The best-fit dependence
is overlaid as the straight line.
}
\label{fig3}
\end{figure}

The $^{152}$Gd $\alpha$-peak at 394~keV served
as a convenient reference standard
to monitor and to correct for
the gain instabilities of the measurement system
in the course of data taking. Accordingly,
data are analyzed in an one-day  interval
and the different data sets  
were added together after calibrating their
$\alpha$-peaks to be at the same energy point.
The measured pulse-height 
variation as a function of temperature
is depicted in Figure~\ref{fig3}. 
The PMT gain variation is typically
less than $\rm{0.1 \% ^o C^{-1}}$,
such the origin of this pulse-height
variation is due to change in the
intrinsic light yield of the crystal.
The best fit value of the temperature dependence is
$\rm{(- 0.78 \pm 0.17) \%~ ^o C^{-1}}$, which
is consistent with the results
of -0.67\%~$\rm{^o C^{-1}}$ at 20~$^o$C
for GSO light yield variation
previously reported~\cite{gsotemp}.

Further
quantitative information can be derived from the
studies of the timing of the events, as well
as from the spectral features.

\subsection{Timing Analysis}
\label{sect::timing}

The timing information is used to probe
the fine structures of the various cascade series,
and the results are displayed in Table~\ref{tabtiming}.

\begin{figure}
{\bf (a)}
\centerline{
\epsfig{file=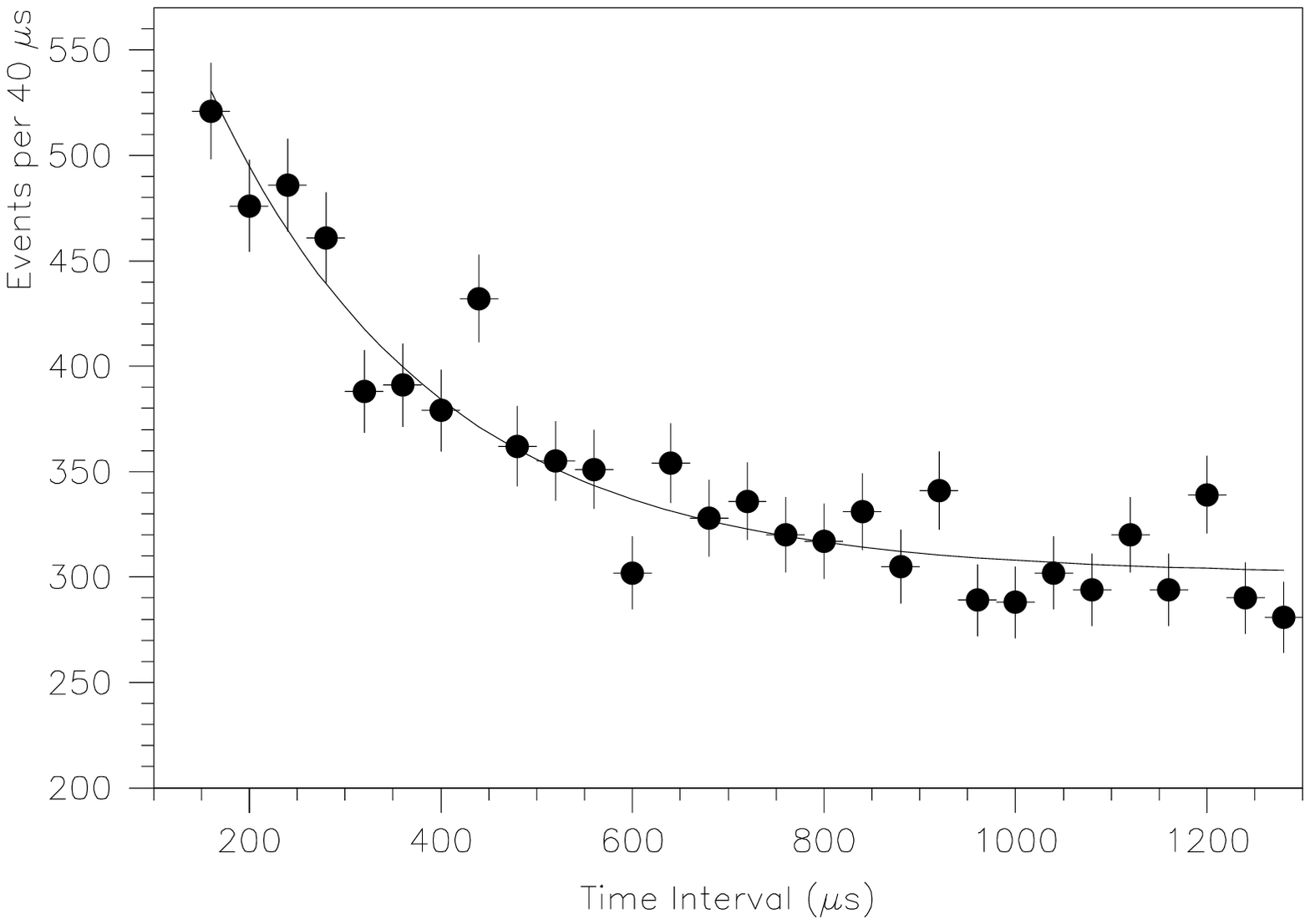,width=9cm}
}
{\bf (b)}
\centerline{
\epsfig{file=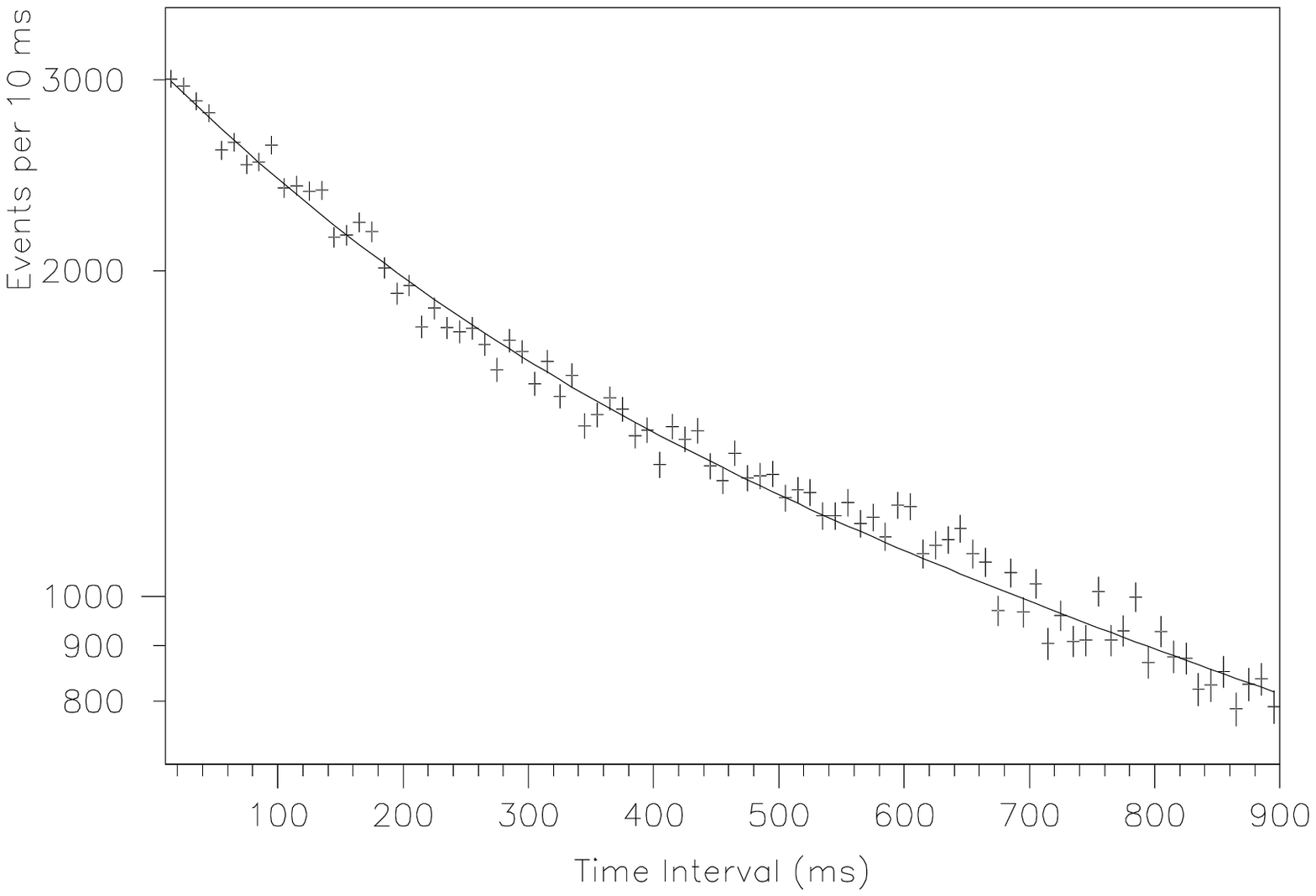,width=9cm}
}
{\bf (c)}
\centerline{
\epsfig{file=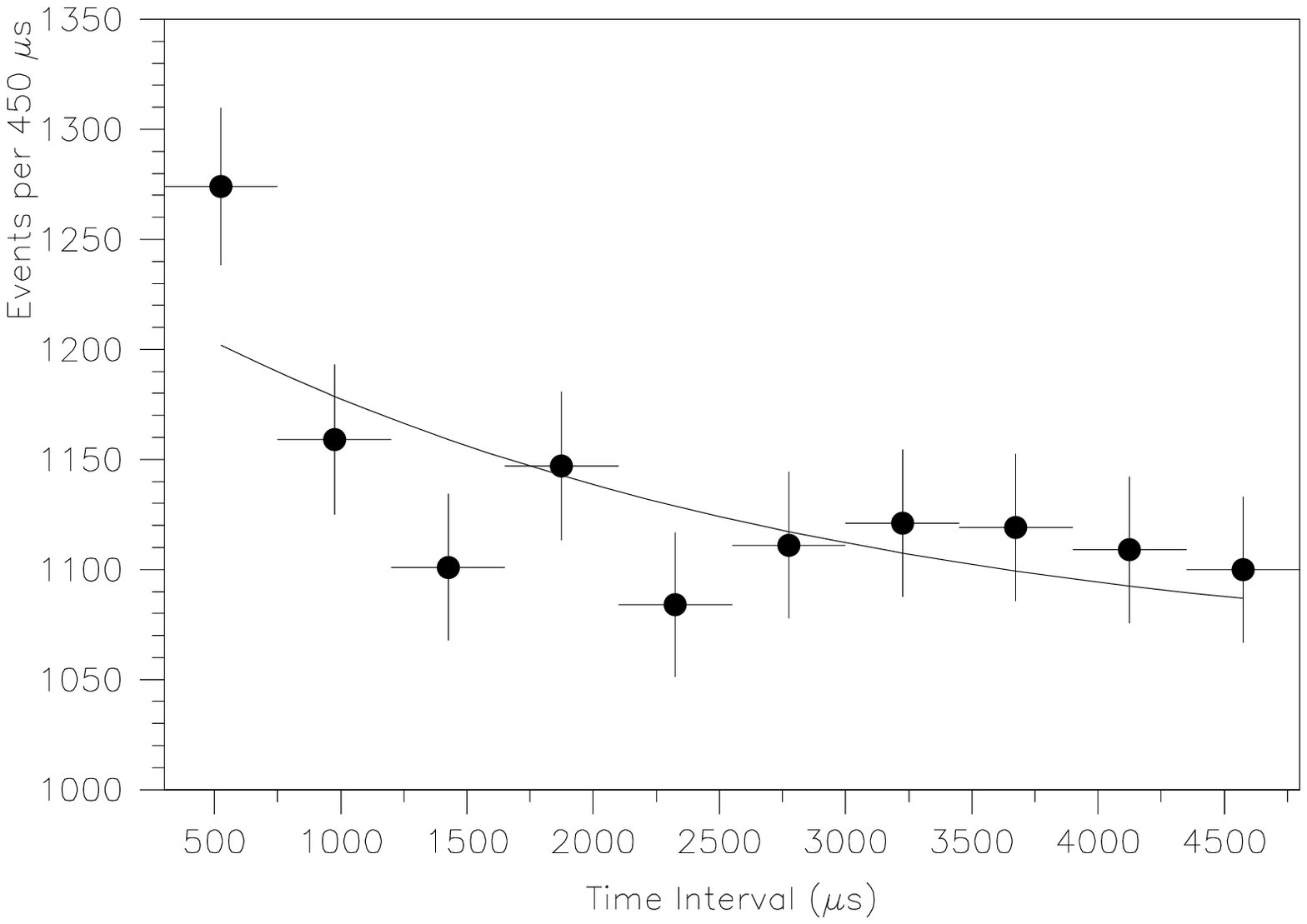,width=9cm}
}
\caption{
Distribution of the time intervals 
between 
(a) the $\beta$
decay of $^{214}$Bi and the following $\alpha$ decay of $^{214}$Po,
from 559~hours of data;
(b) the $\alpha$
decay of $^{220}$Rn and the subsequent $\alpha$ decay of $^{216}$Po,
from 48~hours of data;
(c) the $\alpha$
decay of $^{219}$Rn and the following $\alpha$ decay of $^{215}$Po,
from 559~hours of data.
The solid lines shows the fits of the data 
as discussed in the text.
}
\label{fig4}
\end{figure}

Timing-correlation analysis was applied to the $\u238$ series
where
the $^{214}$Bi-$^{214}$Po decay sequence with
time-correlated $\beta$-$\alpha$ signatures were
selected:
\begin{displaymath}
\rm{
^{214}Bi ~ \rightarrow ~ ^{214}Po ~ + ~ \nuebar ~ +
~ e^- ~ + ~ \gamma 's ~  (Q=3.28~MeV ~ ; ~ T_{\frac{1}{2}}=19.8~min) ~,
}
\end{displaymath}
\begin{displaymath}
\rm{
^{214}Po ~ \rightarrow ~ ^{210}Pb~ + ~ \alpha ~
(Q=7.69~MeV ~ ; ~ T_{\frac{1}{2}}=164~ \mu s) ~~~.
}
\end{displaymath}
Since the $\beta$-decay energy is continuous up to the
end-point energy, the initial events were
selected with energy
from 0.1~MeV to 3.3~MeV, while the second was
within 3$\rm{\sigma_E}$ of the mono-energetic
$\alpha$-energy. 
An exponential
function at nominal decay half-life
plus a constant background 
was employed in the fit. The inclusion
of a flat background was to account for accidental coincidences which become
significant when the sought impurity concentration is tiny. The results
are shown in Figure~\ref{fig4}a and summarized in Table~\ref{tabtiming}. 
The measured activity is $(1.34 \pm 0.22)$~mBq/kg.

In the $\th232$  series,
the
$^{220}$Rn-$^{216}$Po decays:
\begin{displaymath}
\rm{
^{220}Rn ~ \rightarrow ~ ^{216}Po~ + ~ \alpha ~
(Q=6.29~MeV ~ ; ~ T_{\frac{1}{2}}=55~s) ~,
}
\end{displaymath}
\begin{displaymath}
\rm{
^{216}Po ~ \rightarrow ~ ^{212}Pb~ + ~ \alpha ~
(Q=6.79~MeV ~ ; ~ T_{\frac{1}{2}}=0.15~s) ~,
}
\end{displaymath}
can be tagged by an $\alpha$-$\alpha$ delayed coincidence.
Pairs of adjacent events which have 
the correct electron-equivalence energies within 
2$\rm{\sigma_E}$ were selected. 
Since the accidental rates in this case are 
comparable to the decay lifetime of the cascade
sequence, the measured time-correlation  
would be faster than the nominal lifetime,
and the background would be a slow exponential
rather than a constant.
The measured distribution of the 
time intervals of such pairs of events
is shown in Figure~\ref{fig4}b. It was
fitted to two exponential functions with variable
lifetimes. 
The slow component gives  a best-fit value of 
$\rm{\tau_{slow} = 1.1~s}$, which is 
a measurement of the 
the inverse of the background rate at this energy window.
The fast component is 	
$\rm{\tau_{fast} = 204~ms}$, which 
can be understood by Poisson statistics considerations,
where
$\rm{
\tau_{fast}^{-1} =  \tau_{signal}^{-1} + Bkg 
= \tau_{signal}^{-1} + \tau_{slow}^{-1} ~ .
}$
This relation
was experimentally checked to within an 
uncertainty of 10\%.
Folding in an over-counting factor (originated, 
for instance, from that
the parent-accidental-daughter sequence would
be counted as more than one event) of 1.2
obtained from simulations,
an activity of $( 96.0 \pm 8.6 )$~mBq/kg 
for this cascade sequence is derived.
Accordingly, contaminations from the $\th232$ 
series are expected to be the dominant
background contribution for the crystal.

Finally, the decay sequence of $^{219}$Rn--$^{215}$Po
in the $^{235}$U series 
with an $\alpha$-$\alpha$ tag 
was likewise processed:
\begin{displaymath}
\rm{
^{219}Rn ~ \rightarrow ~ ^{215}Po~ + ~ \alpha ~
(Q=6.82~MeV ~ ; ~ T_{\frac{1}{2}}=3.96~s) ~,
}
\end{displaymath}
\begin{displaymath}
\rm{
^{215}Po ~ \rightarrow ~ ^{211}Pb~ + ~ \alpha ~
(Q=6.39~MeV ~ ; ~ T_{\frac{1}{2}}=1.78~ms) ~~~.
}
\end{displaymath}
Events with the corrected sequence of energies were
selected.
The fit to the data with
an exponential decay at nominal half-life
plus a constant background
is shown in Figure~\ref{fig4}c.
The summary of the results is given in Table~\ref{tabtiming}.

The natural abundance of $^{235}$U in uranium is 0.7\%. 
However, the results based on this analysis shown in 
Table~\ref{tabtiming} implies a $^{235}$U/$^{238}$U ratio
to be 4\%.
This inconsistency suggests that the assumption of having
the decay series in secular equilibrium in the crystal
is not valid.

\subsection{Energy Response to $\gamma$ and $\alpha$}
\label{sect::response}

The energy calibration to photons and electrons
was achieved by the standard $\gamma$-ray sources
($^{133}$Ba, $^{22}$Na, $^{137}$Cs, $^{54}$Mn). 
This was fitted to a response function
\begin{equation}
\label{eq::reso}
\rm{
\Bigl({\sigma_E \over E}\Bigr)^2 = \Bigl({A \over \sqrt{E}}\Bigr)^2 +
  \Bigl({B \over E}\Bigr)^2 ~~~
}
\end{equation}
where the energy E is in MeV.
The two terms originate from
the contributions of photo-electron statistics 
and electronic noise, respectively. 
The calibration measurements 
gave best-fit coefficients of A=$9.4 \pm 1.4$\% 
and B=$1.7 \pm 0.7$\%,
which were adopted in subsequent analysis.
The energy resolutions are 12.4\%, 8.5\% and 7.2\% for 0.66~MeV,
1.27~MeV and 1.73~MeV, respectively.

Different particles (electrons, protons, $\alpha$, ...),
give rise to events with different ionization densities dE/dx.
High charge densities will lead to reduced
light yield or ``quenching'', due to recombinations and
other non-radiative dissipation. Accordingly, heavily
ionizing particles like $\alpha$'s  will give less light
than electrons at the same kinetic energy deposited
at the scintillator.

The response of the crystal scintillator to $\alpha$-particles
is very important to evaluate the intrinsic radio-impurities
background.
The quenching factor of the $\alpha$'s depends on the
energy and can be 
parametrized~\cite{scinbasic} as:
\begin{equation}
\label{eq::quen}
\rm{
\epsilon_\alpha = {a\cdot E_\alpha \over 1 + b\cdot E_\alpha} ~~~ ,
}
\end{equation}
where $\rm{E_\alpha}$ and $\epsilon_{\alpha}$ are
the initial kinetic energy of the $\alpha$ and its
``electron-equivalence'' light yield, respectively.
Based on the peak positions of the $\alpha$-particles
from $^{152}$Gd and $^{241}$Am ($\rm{E_\alpha = 2.14~and~
5.4~MeV,~respectively}$),
the values of
$\rm{a=0.19~MeV^{-1}}$ and 
$\rm{b=0.009~MeV^{-1}}$ were derived and adopted
in the global fitting procedures discussed in
Section~\ref{sect::spectrum}.
As an illustration, $\alpha$-particles with
6~MeV kinetic energy, typical for decays
in the $\u238$ and $\th232$ series,
produce an electron-equivalence light
output of 1.08~MeV.

\subsection{Spectral Shape Analysis}
\label{sect::spectrum}

A detailed Monte Carlo simulation on the
contributions of
intrinsic radio-impurities  to the 
background spectrum was performed.
The simulation software was based on the 
EGS4 package~\cite{egs4}.
The electron and photon interaction
cross sections with a GSO crystal were created by specifying
the correct composition of 
2:1:5 for the Gd:Si:O ratio and the exact
geometry of the crystal.
Electrons and photons were traced down to the energy of 10 keV
before discarding.
The number of events generated is of the same range as that recorded
in the data taking time of 1656~hours, such that the statistical
uncertainties of the simulated and measured
spectra are comparable.

An event generator was developed
incorporating all the decay schemes based on 
the standard values~\cite{isotab} in
the $^{232}$Th and $^{238}$U series. 
The minor ($< 10^{-3}$) contributions from
the $^{235}$U series were not included.
The sources were randomly and uniformly located in the
crystal volume, emitting $\alpha$, $\beta$,
and $\gamma$ particles at arbitrary directions. 
The different responses to photons/electrons
and to $\alpha$'s were incorporated based
on the equations
in Section~\ref{sect::response}.
While $\alpha$ and $\beta$ emissions 
deposit all their kinetic energies in the crystal, 
the self-attenuation to the $\gamma$'s by
the crystal are accounted for by simulations.
The combined effects  of short-lived ($\rm{< 1 ~ \mu s}$)
cascades were generated by convoluting the various 
decay channels and their energy depositions.

A global fit was performed, adopting as
free parameters 
the unknown activities of all the long-lived
precursors of the $\th232$ and $\u238$ series,
plus that of the $\alpha$-activities from $^{152}$Gd.
The $\th232$ series has three families
with long-lived precursors, as shown in Table~\ref{tabfitthandu}:
$^{232}$Th and  $^{228}$Ra are unconstrained,
while  activities of $^{228}$Th and its
daughters are constrained by the measurements
due to timing analysis given in Section~\ref{sect::timing}.
Similarly, the $\u238$ series has five families, 
as shown in Table~\ref{tabfitthandu}:
$^{238}$U, $^{234}$U, $^{230}$Th and $^{210}$Pb
are unconstrained,
while activities of $^{226}$Ra and its daughters 
are bounded within the measurement values in
Section~\ref{sect::timing}.

The energy spectra due to
each of these precursors and their daughters
were simulated at the nominal
resolution and quenching parameters based
on Equations~\ref{eq::reso}
and \ref{eq::quen}, respectively,
in Section~\ref{sect::response}.
The global fit was performed by varying the
relative activities between the families
to match the overall spectral shape, within
the energy range of 300~keV to 1.9~MeV.
Events below this range are subjected to
instabilities in electronic noise 
and trigger conditions
during data taking, while those with energies
above 2~MeV are mostly due to ambient radioactivity.
Statistical uncertainties for both the measured 
and simulated spectra were taken into account.

\begin{figure}
\centerline{
\epsfig{file=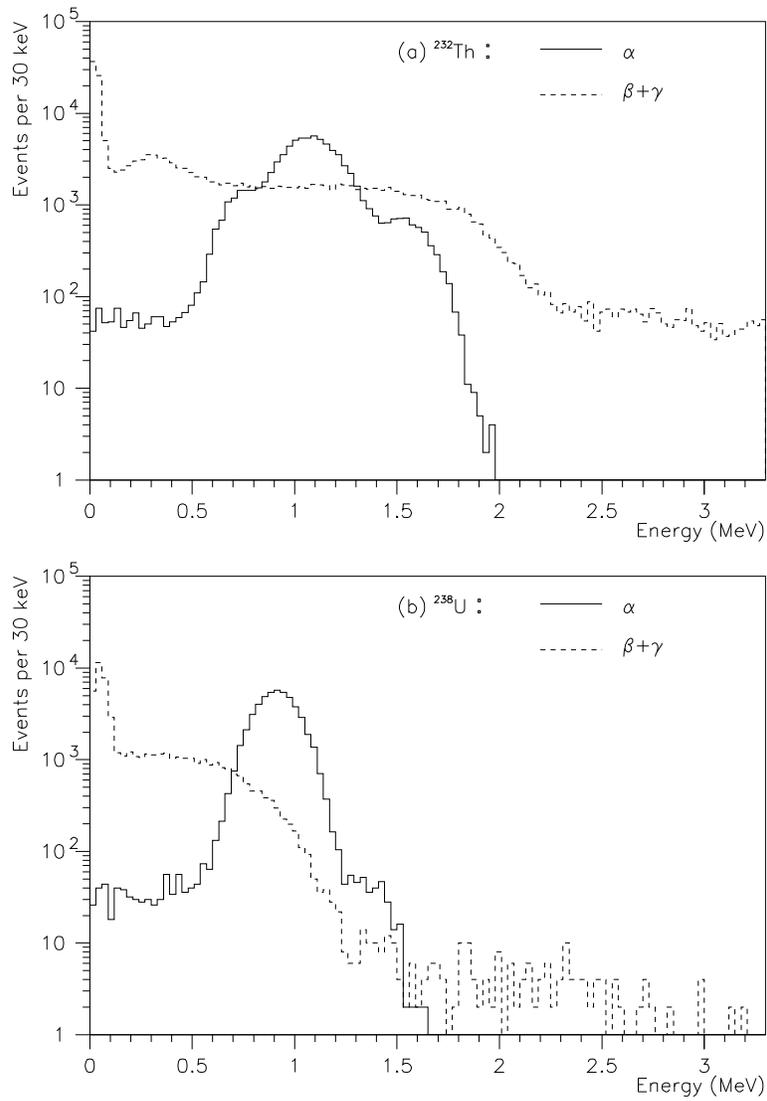,width=12cm}
}
\caption{
Simulated energy spectra due to  the 
$\alpha$ and $\beta + \gamma$ contributions
from the (a) $\th232$ and (b) $\u238$ series.
}
\label{fig5}
\end{figure}

The best fit spectrum 
is displayed alongside with the
measured spectrum in 
Figure~\ref{fig2}. 
An illustration to the contributions
from individual components to the
spectral shape is
shown in Figure~\ref{fig5}:
$\alpha$'s and $\beta$+$\gamma$'s
for the (a) $\th232$ and 
(b) $\u238$ series, respectively.
The measured activity for
$^{152}$Gd $\alpha$-activity (2.9~Hz in this crystal)
corresponds to an isotopic abundance of 0.29\%,
larger than the nominal values of 0.2\%.
An alternative explanation for the enhancement
of the $\alpha$-peak at the electron-equivalence
energy of 0.39~MeV can be the presence of
$^{147}$Sm contamination at the $8 \times 10^{-7}$~g/g level. 
This isotope decays by $\alpha$-emission
at an energy of 2.23~MeV and a half-life of 
$1.06 \times 10^{11}$~yr. 
The two cases, however, are experimentally indistinguishable.

The activities for the families within the
$\th232$ and $\u238$ series 
are shown in Table~\ref{tabfitthandu} and Table~\ref{tabfitthandu},
respectively. 
There are two ambiguities in the exact interpretations
for the physical processes but the difference do not have
big effects on the practical concern of accurate
background subtraction:
\begin{description}
\item[I.]
There are three ``lone'' $\alpha$-emissions from
precursors ($\th232$, $^{234}$U, $^{230}$Th)
which are the only member within their
families. Since they are not accompanied by
other decays and their energies are equal
within $\rm{1.7~\sigma_E}$, the constraints 
from the fits are weak, indicating that
the signatures are experimentally equivalent. 
Only a total activity of 217~mBq/kg can 
be derived.  
However, assessing from
the observations that the crystal is 
dominantly contaminated by the other isotopes
further down the $\th232$ series, it
can be expected that much of the activity
can be attributed to the $\alpha$-decays
of $\th232$.
The component spectra in Figure~\ref{fig5}
assumes equal sharing among the three
isotopes for illustration purpose.
\item[II.]
Similarly, in the energy range of interest,
the $\u238$ family gives two
$\beta$-decays with end-point of 2.2~MeV
and an $\alpha$-decay at 4.27~MeV.
These can in principle mimic the lone $\beta$'s from
$^{228}$Ac and the lone $\alpha$'s discussed
above, both of which are at the same energies.
Again, since the isotopes further down
the $\th232$ series are known to be the dominant
sources, the entire $\beta$-strength
was assigned to the $^{228}$Ac decays
in Table~\ref{tabfitthandu} and Figure~\ref{fig5}.
\end{description}

It can be seen that the radio-isotopes within the
two series
are {\it not} in exact secular equilibrium. 
Nevertheless, a consistent picture can be drawn from
the data where the $\th232$ series dominate the
background at a 100-1000~mBq/kg level, while 
the $\u238$ series
are much less at the $<$10~mBq/kg range $-$
except for $^{210}$Pb, the only long-lived daughter
isotopes from
radon contamination, which is at an activity
level of 200~mBq/kg.

Events below 100~keV 
originate from the $\beta$-decay of $^{228}$Ra 
and $^{210}$Pb.
The peak at 390~keV is due to $\alpha$-decays of $^{152}$Gd.
The various $\alpha$-decays in both the $\th232$ 
and $\u238$ series
populate the broad peak around 1 MeV,
defining the higher and lower energy edges,
respectively.
The shoulder between
1.4 and 2.0 MeV is due to the $\beta - \gamma$ cascades
from $^{228}$Ac and $^{212}$Bi
in the $\th232$ series.
The higher energy events above 2.0 MeV
comes from (a) the $\beta$-decay with
$\gamma$-emissions of $^{208}$Tl, 
which has a decay Q-value of 5.0~MeV, and (b)
the convoluted energy of $\beta + \alpha$
within the 600~ns integration-time
in the $^{212}$Bi$-$$^{212}$Po 
cascade. 
The intermediate $^{212}$Po half-life 
is short (299~ns) so that 75\% of decay-sequence
are  combined together as a single event.

The relative residual  
(that is, [Measured$-$Simulated]/Measured) 
as a function of energy 
is depicted in Figure~\ref{fig6}a.
Only statistical uncertainties are shown.
It can be seen that, from 300~keV to 1.9~MeV
where the fit was performed,
the simulated spectrum matches
well to the measured background to
better than the few \% level, except in
the ``trough'' region between the two peaks
at about 600~keV. 
The measured spectrum between 700~keV and 1.9~MeV,
which is dominated by the $\alpha$ and $\beta$ events
for both $\th232$ and $\u238$ components
and relevant for double
beta decay analysis discussed
in Section~\ref{sect::dbd},
is fitted well
by the simulated spectra.
Among them, the $\alpha$-dominated
bump (700~keV to 1.3~MeV) gives larger
dispersions in their residuals. 
This can be understood by the
fact that the detailed spectral shape in
this region is due to the 
convolutions of many $\alpha$-lines 
folded in with their respective quenching factors.
In comparison, the shape in the 1.4~MeV to 
1.9~MeV range is due to
a high energy $\alpha$-decay (with energy 8.78~MeV
from $^{212}$Po)
then a $\beta$-decay (Q-value 2.13~MeV from 
$^{228}$Ac), 
leading to more accurate and robust matching.
The big excess above 2~MeV over
contributions from intrinsic radioactivity
confirms that ambient background,
due to the $\gamma$-rays from $^{208}$Tl,
are dominant in this region.
It can be understood from that 
low energy ambient photons suffer larger
suppression factors due to the NaI(Tl)  
anti-Compton detector.

\subsection{Double Beta Decay Analysis}
\label{sect::dbd}

The isotope
$^{160}$Gd is a $\beta\beta$ emitter at a Q-value of
1.73 MeV. The $\beta\beta$ lifetime limits
from the present GSO measurements 
are not as sensitive as the various other
double beta decay experiments~\cite{bbreview},
such as the best sensitivities achieved
in $^{76}$Ge with high-purity germanium
detectors~\cite{hpge}. The analysis was
nevertheless performed for completeness and
for studying how the background subtraction
results can be applied in a realistic case.
Different approaches for setting the limits
are compared.

\begin{figure}
{\bf (a)}
\centerline{
\epsfig{file=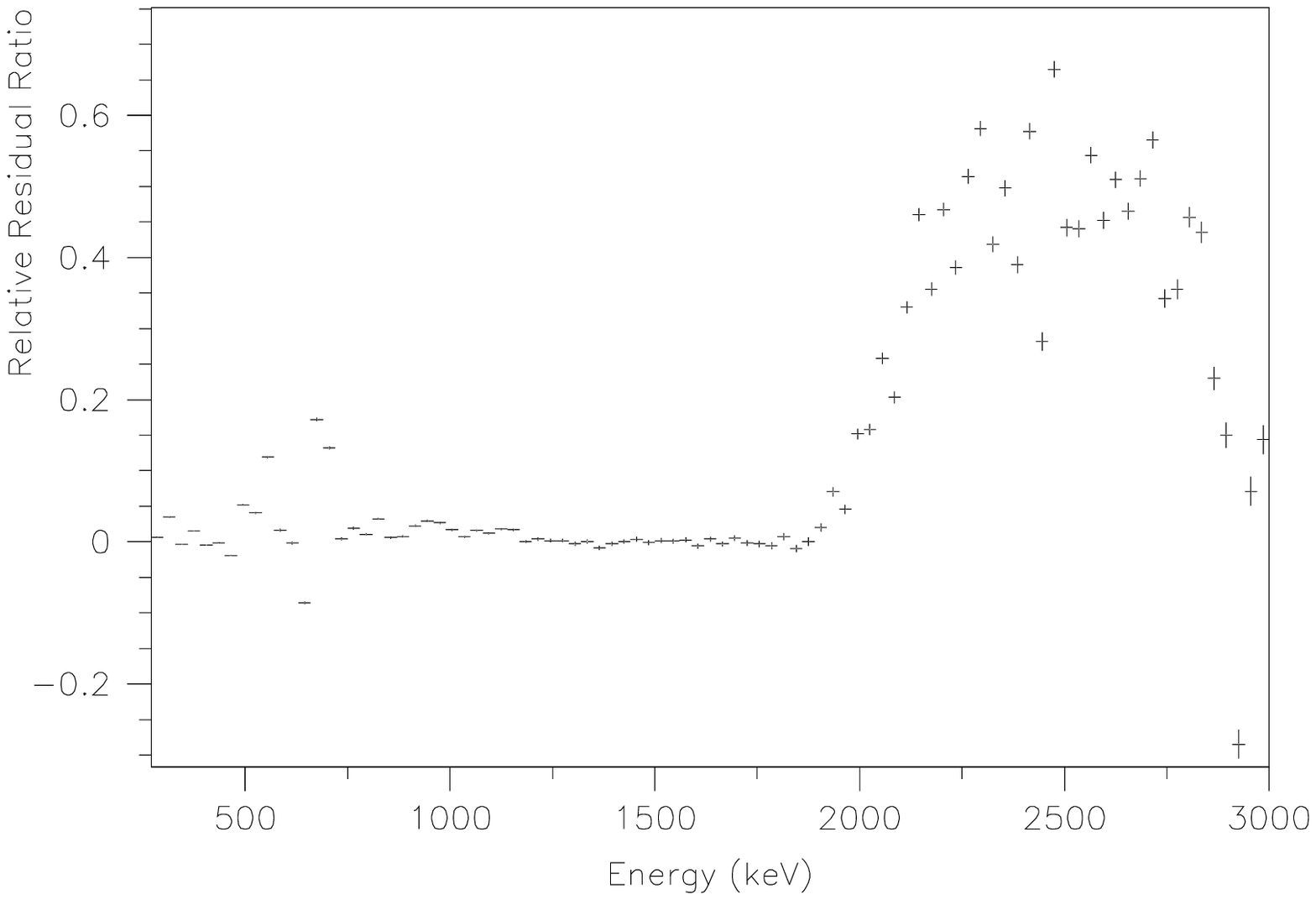,width=12cm}
}
{\bf (b)}
\centerline{
\epsfig{file=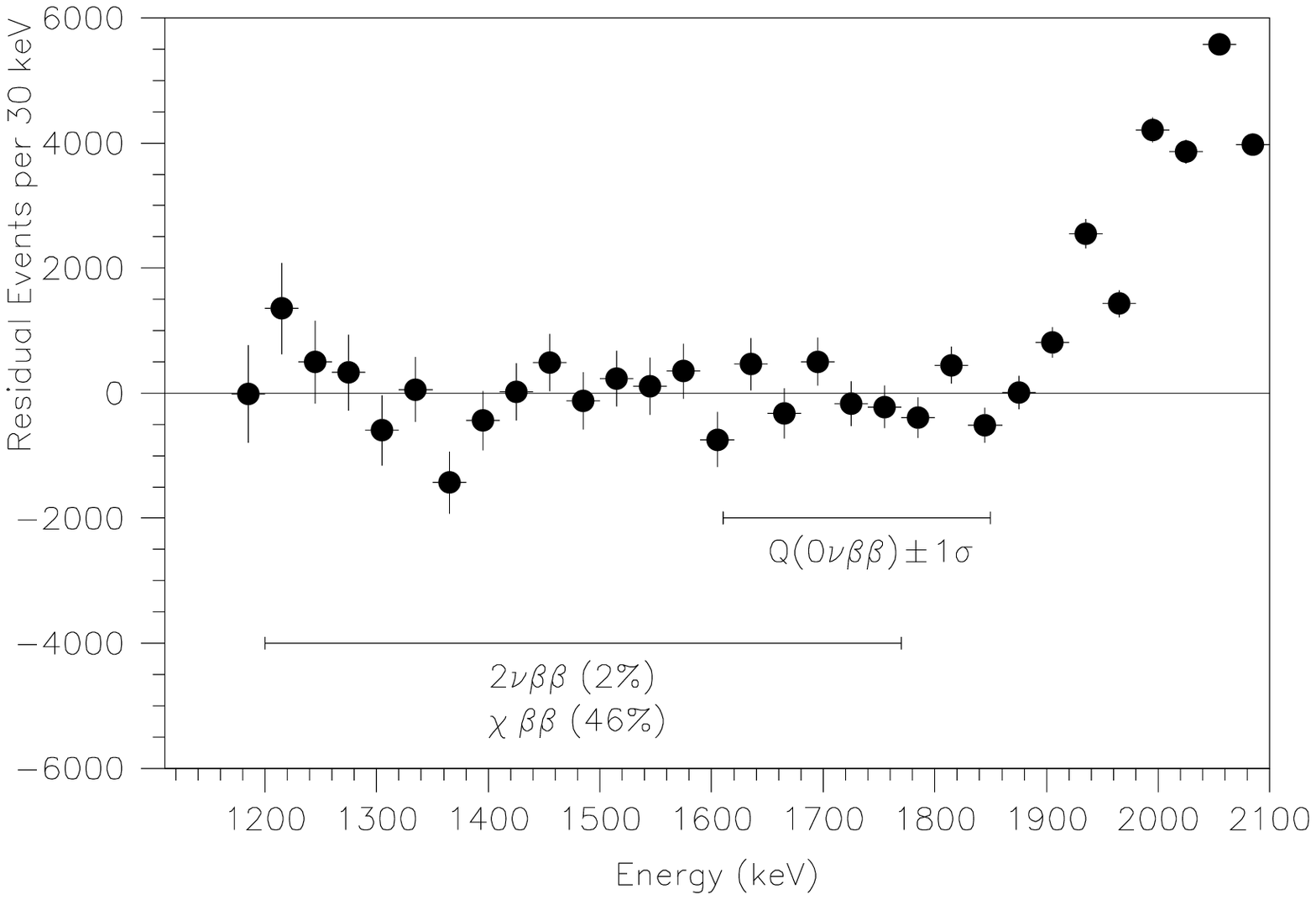,width=12cm}
}
\caption{
(a)
Ratio of the residual (measured$-$simulated)
to the measured background 
spectrum for 1656~hours of data.
(b)
Residual events at the energy range 
relevant for double
beta decay analysis. 
Statistical error bars are shown.
}
\label{fig6}
\end{figure}

The measured raw background
rate in the present shielding and
veto configuration in the vicinity of the end-point
is 1~$\rm{kg ^{-1} keV ^{-1} hr ^{-1}}$.
From the results in Section~\ref{sect::results},
all of these events
can be accounted for to be due to
intrinsic radioactivity, mostly
$\beta$-decays, at an
uncertainty level of better than
a few \%.
The contributions from these known sources
were subtracted from the raw
spectrum based on 1656~hours of data,
and the residual spectra around
the $0 \nu \beta \beta$ end-point 
is shown in
Figure~\ref{fig6}b, showing
no particular structures.
Statistical error bars were adopted.
A $\chi^2$ per degree of freedom ($\chi^2$/dof)
of 41.9/25 was obtained
for a ``zero residual background''  hypothesis
between 1.2~MeV to 1.9~MeV.
The best-fit amplitude for
a possible Gaussian signal
at the expected resolution 
above zero-background
at the 1.73~MeV end-point
is $\rm{66 \pm 1346}$.
Assuming Gaussian errors,
the upper limits for the signal ($\rm{S_u}$)
are
$\rm{
S^{0 \nu \beta \beta}_u < 699(1775) ~ at ~  68(90)\%~CL.
}$

It is illustrative to compare this with the other values
of $\rm{S^{0 \nu \beta \beta}_u}$ derived by alternative methods,
which are based on the {\it assumption} that the
background is a smooth spectrum at the 
end-point region, without
prior knowledge of the actual 
background components~\cite{bbgso2}.
There are 671545 events observed
within the $\pm 1~\sigma$ (efficiency 0.68)
region around the end-point. 
{\it If} these are all due
to background, the signal must be smaller than
the statistical fluctuations such that, folding
in the efficiency factor,\\
\hspace*{3cm}
$\rm{ S_u (90\%~CL) < ( 1.28 \times \sqrt{Counts} )/ 0.68 =  1543.}$\\
Alternatively, the spectrum can be fitted to
an {\it ad hoc} polynomial and the residual
at the $\pm 1~\sigma$ region is
$\rm{39 \pm 1159}$, giving 
$\rm{ S_u (90\%~CL) < 2239~.}$
Therefore, it can be seen that the background
identification and subtraction procedures adopted
in this work gives similar performance to
subtraction schemes based on unknown
background source and assumed spectral
shape. The present scheme provides further
justifications and
robustness to the background suppression,
and is applicable for signals other than
the easily distinguishable energy peaks (such
as continuous spectra)
as well as at regions where the background
spectra do exhibit certain structures
(indeed, there is a bump at 1.6~MeV due to a
high energy $\alpha$-decay).
The fact that the present uncertainty is 
comparable to the 
statistical fluctuations 
($\rm{\sqrt{Counts}}$)
indicates that, up to a suppression factor
of $10^{-3}$, 
the procedure is 
still statistics-limited and there
are room to enhance the sensitivities further
with more data and longer simulation time.

The half-life limit can be expressed as  
\begin{equation}
\rm{
T^{0 \nu \beta \beta}_{\frac{1}{2}} ~ > ~ log_e 2  \times 
\frac{ time \times N ( ^{160}Gd )  }{S^{0 \nu \beta \beta}_u} ~~.
}
\end{equation}
Putting the experimental values of 
the target number 
$\rm{N ( ^{160}Gd ) = 1.07 \times 10^{24}}$,
and the data taking time of 
$\rm{1656~hours (0.19~yr)}$
and the values of $\rm{S_u}$ derived
above, we obtain
\begin{displaymath}
\rm{
T^{0 \nu \beta \beta}_{\frac{1}{2}} ~ > ~
2.0 (0.8) \times 10^{20}~y  
~~~ at ~~~   68(90)\%~C.L.~,
}
\end{displaymath}
consistent with and at a comparable range 
to the previous results:
$\rm{T_{\frac{1}{2}} > 
3.0  \times 10^{20}~y}$ at 68\%~C.L.~\cite{bbgso2}
and 
$\rm{T_{\frac{1}{2}} > 
8.2  \times 10^{20}~y}$ at 90\%~C.L.~\cite{bbgso3}.
These limits apply to both the
$0^+ \rightarrow 0^+$ ground state
to ground state transition
as well as the $\rm{0^+ \rightarrow 2^+ (87~keV)}$
transition to the excited state, since
the 87~keV $\gamma$-rays emitted
are fully absorbed by the crystal.
The corresponding limits of the effective neutrino mass  parameter,
based on calculations in Ref.~\cite{bb2m}, are
$ \rm{
\langle m_{\nu}  \rangle  <   65(103) ~ eV 
} $ at 68(90)\% CL.

Limits can also be set for the 
two-neutrino double beta decay (${2 \nu \beta \beta}$) 
and the neutrinoless double beta decay with Majoron emission
(${\chi \beta \beta}$) channels.
We took the conservative approach of deriving
the limits only with the residual data in Figure~\ref{fig6}b
between the 1.2~MeV and 1.9~MeV region.
The $\chi^2$/dof
is good for a zero-residual hypothesis
using statistical error bars alone, indicating that
the background in this region is well-understood.
The efficiency factors for the ${2 \nu \beta \beta}$ and
${\chi \beta \beta}$ channels to be within this energy range
are 2\% and 46\%, respectively.

The best-fit of a ${2 \nu \beta \beta}$ spectra over zero-residual
in this energy range gives the optimal activity of
$(34620 \pm 67167)$ events, from which
$\rm{
S^{2 \nu \beta \beta}_u <  66188(120593) ~ at ~  68(90)\%~CL,
}$ leading to 
\begin{displaymath}
\rm{
T^{2 \nu \beta \beta}_{\frac{1}{2}} ~ > ~
2.2 (1.2) \times 10^{18}~y  
~~~ at ~~~   68(90)\%~C.L.
}
\end{displaymath}
Similarly, a fit to the $\chi \beta \beta$ spectra
gives the best value of 
$(-1092 \pm 4138)$ events. Restricting a possible signal to
be positive definite, the upper limits 
$\rm{
S^{\chi \beta \beta}_u <  4676(7241) ~ at ~  68(90)\%~CL,
}$ is derived, giving rise to
\begin{displaymath}
\rm{
T^{ \chi \beta \beta}_{\frac{1}{2}} ~ > ~
3.0 (1.9) \times 10^{19}~y  
~~~ at ~~~   68(90)\%~C.L.
}
\end{displaymath}
Though these $2 \nu \beta \beta$ and $\chi \beta \beta$ limits
are only modest compared to those of the other double beta decay
candidates, they represent nevertheless more than
an order of magnitude
improvement over previous published results 
on $^{160}$Gd~\cite{bbgso1}.
The improved sensitivities can be attributed to
that the background must first be quantitatively
understood and the detailed spectral structures
accounted for to allow them be subtracted
off from signals which are themselves continuous
spectra without specific structures.

\section{Discussions and Summary}
\label{sect::outlook}

The analysis methods and results discussed in
this article give indications of the potentialities
and problems 
of GSO crystals in neutrino experiments. They also
provide training grounds in developing
the schemes for the 
background understanding and suppression
for similar class of experiments.

Low background experimentation techniques must
be used to maximize the sensitivities in a realistic
experiment.
To get to the relevant $10^{24}$~y sensitivity range 
for double beta decay searches,
one needs $10^{8}$ improvement in the [(target mass$\times$time)/background]
value. With a realistically achieve-able target mass of O(10~ton) 
and running time of O(1~year), the background has to be further reduced
by at least
factor of 10$^3$, via more elaborate shielding configurations
and data taking at an underground laboratory, as well
as through careful choice of  raw materials  
with reduced $\th232$ and $\u238$ activities,
especially the contaminations due to $^{228}$Ra.
Measurements with raw background rate of 
0.04~$\rm{kg ^{-1} keV ^{-1} hr ^{-1}}$
at the $0 \nu \beta \beta$ end-point were
reported~\cite{bbgso3}, such that factor of hundreds
in improvements have still to be worked on.
An alternative or perhaps better candidate for
double beta decay experiments with 
crystal scintillators $^{116}$Cd
with cadmium tungstate (CdWO$_4$) crystals~\cite{cdwo4}.

Solar neutrino experiments with $^{160}$Gd typically
involve an event 
rate of O(0.1~ton$^{-1}$day$^{-1}$),
and hence imply more stringent intrinsic radio-purity 
level requirements. 
The specific delayed-correlation
signatures discussed in Section~\ref{sect::intro}
provides powerful means to suppress
background due to accidental coincidence.
Design studies~\cite{lens} indicate that
the purity levels of $< 10^{-10}$~g/g are
required for $^{238}$U and $^{232}$Th series
and $10^{-15}$~g/g for $^{235}$U (where $\beta$-decays
of $^{231}$Th with a delayed
$\gamma$-emissions may mimic the neutrino signatures). 
The results in Table~\ref{tabtiming} show that improvements 
to the $10^3$ and $10$ levels 
relative to the activities
measured in this work for the $\th232$ and $\u238$ series, 
respectively, are necessary. 
This timing analysis results in Table~\ref{tabtiming}, 
however, 
only gives the contamination levels of 
$^{227}$Ac and its short-lived daughters within
the $^{235}$U series but 
does not provide direct measurement
of the $^{235}$U concentration.
If secular equilibrium is assumed, an improvement
of better than $10^3$ is necessary.
The electron-equivalence light output of 392~keV
for $^{152}$Gd $\alpha$-decay is in between the
expected range of $\rm{\nu_e}$(pp) and $\rm{\nu_e}$($^7$Be).
Effects of the tails  of the energy distribution 
to the $< 10^{-3}$ level have to be studied.
There are isotope-separation
schemes to deplete $^{152}$Gd~\cite{snugso},
thereby reducing this potential background.
In addition, care must be exercised in selecting the
sources of cerium the
dopant materials in order to avoid contaminations
from its long-life isotopes:  
$^{139}$Ce ($\rm{T_{\frac{1}{2}} = 140~days}$)
which decays by electron  capture emitting $\gamma$-rays
of 166~keV, and $^{144}$Ce ($\rm{T_{\frac{1}{2}} = 284~days}$)
which decays by $\beta$-emissions at a
Q-value of 320~keV.

A GSO scintillating crystal detector which can 
satisfy the requirements for solar neutrino experiments
will also be a powerful device
for double beta decay searches that can operate simultaneously.

One of the problems for the construction of big detector 
systems based on GSO crystals has been the cost.
There are ongoing efforts from the industries to produce
and market this crystal at a similar price range as BGO~\cite{sun}.
If achieved, this crystal will be an attraction option
for electromagnetic calorimeter in a high-rate environment
as well as for medical imaging applications.

An independent motivation of this work is  
to study the methods for
the measurements and subtraction
of background due to intrinsic radio-isotopes.
In a scintillating crystal detector with minimal passive
volume and where a three-dimensional
fiducial volume definition can be achieved,
the ambient background in the sub-MeV range
will be highly reduced
and the primary experimental focus will be one on
the understanding and suppression of internal
background~\cite{prospects}. 
In addition, forthcoming big liquid scintillator
experiments like BOREXINO~\cite{borexino} and 
KamLAND~\cite{kamland} for
low energy solar neutrino and long-baseline
reactor neutrino experiments will 
need to identify and measure the background
processes quantitatively to justify their
signals, in a similar spirit to this work.
The GSO crystal used in this measurement,
having relatively high contamination levels,
provides a convenient platform to develop the
analysis procedures and to understand the
various systematics involved. 
Crystals with much better radio-purity level
are available. For instance, the secular
equilibrium levels for the $\th232$ and $^{238}$U series
in CsI(Tl) are below the $10^{-12}$~g/g range~\cite{expt}.

It has been shown that
a subtraction factor of $10^{-3}$ is possible,
at least in the case where the region 
is dominated by a  smooth 
$\beta$-decay background.
The background of $\th232$ and $\u238$ in a
GSO crystal treated in this work
are in fact more difficult to handle from the
background subtraction point of view. There are
multiple $\alpha$-decays to be convoluted together
with their quenching factors
to be taken into account, 
detection efficiency factors for the $\gamma$-emissions
which depends on the details in detector geometry,
as well as convolutions
of various signals due to short life-time cascades.
Other simpler decay schemes are expected to be
easier to handle. For instance, the background
due to $\beta$-decays of $^{40}$K can be more
accurately accounted for, since these
events are uncorrelated with the
other processes such that
the background spectral shape will
be simply the well-known $\beta$-spectra. 
Moreover, the activity can be independently obtained, 
besides being derived from the best-fit value, through 
the measurements of the related 1.46~MeV 
$\gamma$-decays of $^{40}$K.
Furthermore, crystals like NaI(Tl) and CsI(Tl)~\cite{expt},
which have excellent pulse shape discrimination properties
for $\gamma$/$\alpha$ separation
and where the $\alpha$-particles suffer less severe quenching,
can provide much better measurements and identifications
for the $\alpha$-decays from the various channels,
providing further
powerful diagnostic tools to probe the structures
within the various cascade series.

The authors are grateful to
T. Murakami for the loan of the GSO crystal, and to
C. Rangacharyulu for discussions on the 
various decay schemes of radio-isotopes.
The work is supported by contracts
NSC88-2112-M-001-007 
and NSC89-2112-M-001-028
from the National Science Council, Taiwan.

\clearpage

\renewcommand{\baselinestretch}{1.2}

\begin{table}
\begin{center}
\large
\begin{tabular}{|l||c|c|c|c|c|}
\hline
%& & & & & \\
Properties & NaI(Tl) & CsI(Tl) & CaF$_2$(Eu) & BGO & GSO \\ \hline \hline
%& & & & &  \\ 
%& & & & & \\
Density & 3.67 & 4.51  & 3.19 & 7.13 & 6.7  \\
%& & & & & \\
Relative Light Yield & 1.00$^{\dagger}$ & 0.45 & 0.50 & 0.15 & 0.2  \\
%Radiation Length (cm) & 1.85 & 2.59 & 1.12 & 1.38  & $\sim$45 & $\sim$45  \\
%& & & & & \\
Emission Peak $\rm{\lambda_m}$ (nm) & 415 & 530 & 435 & 480 & 440  \\
%& & & & & \\
Decay Time (ns) & 230 & 1000 & 940 & 300 & 30-60 \\
%& & & & & \\
Refractive index at $\rm{\lambda_m}$ 
& 1.85 & 1.80 & 1.44 & 2.15 & 1.85  \\
%& & & & & \\
Hygroscopic & yes & slightly & no & no & no  \\
%& & & & & \\
Melting point ($^o$C) & 651 & 621 & 1418 & 1050 & 1900 \\ \hline
%& & & & &  \\ \hline
%\multicolumn{6}{l}{ }\\
\multicolumn{6}{l}{$^{\dagger}$ Typical light yield for NaI(Tl) 
is about 40000 photons per MeV.}
\end{tabular}
\end{center}
\caption{Characteristic properties of several common
crystal scintillators and their comparison with
the GSO crystals}
\label{scincrystal}
\end{table}

\clearpage

\begin{table}
\begin{center}
\large
\begin{tabular}{|l||c|c|c|}
\hline
Decay series precursor &  $^{232}$Th & $^{238}$U 
& $^{235}$U \\ \hline \hline
Actual decay sequence & $^{220}$Rn$- ^{216}$Po 
& $^{214}$Bi$- ^{214}$Po & $^{219}$Rn$- ^{215}$Po \\
Signatures & $\alpha - \alpha$ & $\beta - \alpha$ & $\alpha - \alpha$ \\
Bounds on time interval (ms)  & 
   [10, 1000]  & [0.13, 1.3] & [0.32, 4.8]  \\
Half-life (ms) &  150 & 0.164 & 1.78 \\
Duration of Data Taking (hours) & 48 & 559 & 559 \\
No. of events  & $28259\pm2543$ & 
$2651\pm445$ & $1015 \pm 588$  \\
Efficiency of Cuts & 
 0.82 & 0.57 & 0.78 \\
Over-counting factors & 1.2 & 1.0 & 1.0 \\
Activity (mBq/kg)  & 
$96.0\pm8.6$   & $1.34\pm0.22$ & $0.37\pm0.21$ \\
Concentration of Precursor (g/g)$^{\dagger}$
& $(2.9\pm0.3)$ 
& $(1.1\pm0.2)$ 
& $(4.4\pm2.5)$ \\ 
& $\times 10^{-8}$ & $\times 10^{-10}$ & $\times 10^{-12}$ \\ \hline
\multicolumn{4}{l}{$^{\dagger}$ Assuming secular equilibrium}
\end{tabular}
\end{center}
\caption{
Summary of results in the determination of the concentration 
of the thorium and uranium series in 
the GSO crystal based on energy-timing correlations. 
The statistical uncertainties are quoted.
}
\label{tabtiming}
\end{table}

\clearpage

\begin{table}
\normalsize
\begin{center}
\begin{tabular}{|l|c|c|c|c|c|}
\hline
Family in & Isotope & Half-Life & Decay Mode & Q-value 
& Activity \\ 
$^{232}$Th series & & & & (MeV) & (mBq/kg) \\ \hline \hline
$^{232}$Th & $^{232}$Th & $1.405\times 10^{10}$ y & $\alpha$
& 4.013 & $^\dagger$ \\ \hline
$^{228}$Ra & $^{228}$Ra & $5.75$ y & $\beta$ &  0.046 & 733~$^{\ddagger}$ \\
& $^{228}$Ac & $6.15$ h &  $\beta$ &  2.127 &   \\ \hline
$^{228}$Th & $^{228}$Th & $1.913$ y & $\alpha$ & 5.423 & 107 \\
 & $^{224}$Ra & $3.66$ d &  $\alpha$ & 5.685 &   \\
 & $^{220}$Rn & $55.6$ s & $\alpha$ & 6.288 &    \\ 
 & $^{216}$Po & $0.145$ s & $\alpha$ & 6.778 &   \\
 & $^{212}$Pb & $10.64$ h & $\beta$ & 0.574 &   \\
 & $^{212}$Bi & $60.55$ min & $\alpha$ (64\%) & 6.090 &   \\
 &           &           & $\beta$ (36\%) & 2.254 & \\
 & $^{212}$Po & $0.299~\mu$s & $\alpha$ & 8.784 &   \\
 & $^{208}$Tl & $3.053$ min & $\beta$ & 5.001 &   \\ 
\hline \hline \hline
Family in & Isotope & Half-Life & Decay Mode & Q-value
& Activity \\
$^{238}$U series & & & & (MeV) & (mBq/kg) \\ \hline \hline
$^{238}$U & $^{238}$U & $4.47\times 10^{9}$ y & $\alpha$
& 4.27 & $<$ 2$~^{\ddagger}$ \\
& $^{234}$Th &  $24.1$ d & $\beta$ &  0.183 &  \\
& $^{234}$Pa$^*$ &  $1.18$ min & $\beta$ (99.87\%)  &  2.207 &  \\
&  &  & IT (0.13\%)  &  0.08 &  \\
& $^{234}$Pa &  $6.70$ h & $\beta$  &  2.287 &  \\ \hline
$^{234}$U & $^{234}$U
   &  $2.45 \times 10^{5}$ y & $\alpha$  &  4.856 & $^{\dagger}$ \\ \hline
$^{230}$Th & $^{230}$Th
   &  $8.0 \times 10^{4}$ y & $\alpha$  &  4.771 & $^{\dagger}$ \\ \hline
$^{226}$Ra & $^{226}$Ra
   &  $1600$ y &  $\alpha$  &  4.871 &  $<$ 2  \\
 & $^{222}$Rn & $3.83$ d &  $\alpha$ & 5.591 &  \\
 & $^{218}$Po & $3.05$ min & $\alpha$ & 6.115 &   \\
 & $^{214}$Pb & $26.8$ min & $\beta$ & 1.024 &  \\
 & $^{214}$Bi & $19.7$ min & $\beta$ & 3.270 & \\
 & $^{214}$Po & $163.7~\mu$s & $\alpha$ & 7.834 &  \\ \hline
$^{210}$Pb & $^{210}$Pb & $22.26$ y & $\beta$ & 0.063 & 198 \\
 & $^{210}$Bi & $5.01$ d & $\beta$ & 1.161 & \\
 & $^{210}$Po & $138.4$ d & $\alpha$ & 5.408 &  \\ \hline
\multicolumn{6}{l}{\small $^{\dagger}$  Experimentally Equivalent.
Total 217~mBq/kg. See Text.}\\
\multicolumn{6}{l}{\small $^{\ddagger}$  Experimentally Equivalent.
Assign weight to $^{228}$Ra.  See Text.}
\end{tabular}
\end{center}
\caption{
Activities of the radio-isotopes in the $\th232$ 
and $\u238$ series,
based on best-fit parameters to the background
spectrum.
}
\label{tabfitthandu}
\end{table}

\clearpage

\clearpage

\pagebreak

\clearpage

\clearpage

\clearpage

\begin{thebibliography}{99}
\bibitem{crystal}
For a recent review on the crystal scintillator detector,
see, for example, \\
M. Ishii and M. Kobayashi,
Prog. Crystal Growth and Charact., {\bf 23}, 245 (1991),
and references therein.
\bibitem{emcalo}
For a recent review on the applications of crystal
scintillator in particle physics, see, for example\\
G. Gratta, H. Newman, and R.Y. Zhu,
Ann. Rev. Nucl. Part. Sci. {\bf 44}, 453 (1994).
\bibitem{prospects}
H.T. Wong et al., 
Astropart. Phys. {\bf 14}, 141 (2000).
\bibitem{dama}
R. Bernabei et al., Phys. Lett. {\bf B 389}, 757 (1996);\\
R. Bernabei et al., Phys. Lett. {\bf B 450}, 448 (1999).
\bibitem{expt}
H.T.~Wong and J. Li,
Mod. Phys. Lett. {\bf A 15}, 2011 (2000);\\
H.B. Li et al., TEXONO Collaboration,
hep-ex/0001001,
Nucl. Instrum. Methods {\bf A 459}, 93 (2001).
\bibitem{bbreview}
For a recent review on double beta decay experiments, see, for example,\\
A. Morales., Nucl. Phys. {\bf B} (Procs. Suppl.) {\bf 77}, 335 (1999).
\bibitem{snureview}
For a recent review on solar neutrino experiments, see, for example,\\
R.E. Lanou Jr., Nucl. Phys. {\bf B} (Procs. Suppl.) {\bf 77}, 55 (1999).
\bibitem{lens}
R.S. Raghavan, Phys. Rev. Lett. {\bf 78}, 3618 (1997);\\
LENS Project, Letter of Intent (1999).
\bibitem{gsofirst}
K. Takagi and T. Fukazawa, Appl. Phys. Lett. {\bf 42}, 43 (1983).
\bibitem{gsoprop}
C.L. Melcher et al.,
IEEE Trans. Nucl. Sci. {\bf 37}, 161 (1990);\\
M. Moszynski et al.,
Nucl. Instrum. Methods. {\bf A 372}, 51 (1996); \\
N. Tsuchida et al.,
Nucl. Instrum. Methods. {\bf A 385}, 290 (1997).
\bibitem{gsodet}
S. Nakayama et al.,
Nucl. Instrum. Methods. {\bf A 404}, 34 (1998); \\
M. Tanaka et al.,
Nucl. Instrum. Methods. {\bf A 404}, 283 (1998).
\bibitem{bbgso1}
S.F. Burachas et al., Phys. Atom. Nucl. {\bf 58}, 153 (1995).
\bibitem{bbgso2}
M. Kobayashi and S. Kobayashi, Nucl. Phys. {\bf A 586}, 457 (1995).
\bibitem{bbgso3}
F.A. Danevich et al., Nucl. Phys. {\bf B} (Proc. Suppl.)
{\bf 48}, 235 (1996).
\bibitem{snugso}
J.-F. Cavaignac, Grenoble group, private communications (1999);\\
M. Nakahata, Tokyo group, private communications (1999).
\bibitem{isotab}
Table of Isotopes, 8th Edition, Vol II, ed. R.B. Firestone et al.,
John Wiley \& Sons, Inc., New York, (1996).
\bibitem{gsotemp}
T. Tsuchida et al.,
Nucl. Instrum. Methods. {\bf A 385}, 290 (1997).
\bibitem{scinbasic}
See, for example,
``Theory and Practice of Scintillation
Counting'',  J.B. Birks, Pergamon (1964).
\bibitem{egs4}
The EGS4 Code System,
W.R. Nelson, H. Hirayama, and D.W.O. Rogers, 
Stanford Linear Accelerator Center Report SLAC-265 (1985).
\bibitem{hpge}
L. Baudis et al., Phys. Rev. Lett. {\bf 83}, 41 (1999).
\bibitem{bb2m}
K. Muto, E. Bender and H.V. Klapdor-Kleingrothaus, 
Europhys. Lett. {\bf 13}, 31 (1990).
\bibitem{cdwo4}
F.A. Danevich et al., Phys. Lett. {\bf B 344}, 72 (1995);\\
S.F. Burachas et al., Nucl. Instrum. Methods 
{\bf A 369}, 164 (1996).
\bibitem{sun}
H.C. Sun, China Institute of Atomic Energy, private
communications (1999).
\bibitem{borexino}
L. Oberauer, Nucl. Phys. {\bf B} (Procs. Suppl.) {\bf 77}, 48 (1999).
\bibitem{kamland}
A. Suzuki, Nucl. Phys. {\bf B} (Procs. Suppl.) {\bf 171}, 48 (1999).
\end{thebibliography}
\end{document}